\documentstyle[12pt,a41,epsfig,wrapfig]{article}

\begin{document}

\begin{center}
{\Large \bf
Realistic arrival time distribution \\from an isotropic light source
} \\

\vspace*{2cm}
{G.~Japaridze$^a$\footnote{
Clark Atlanta University, 223 James P. Brawley Dr., SW, Atlanta, GA 30314, U.S.A. Phone:  +404 880 6420,
Email: japar@ctsps.cau.edu} and  M.~Ribordy$^{b}$\footnote{
Universit\'e de Mons-Hainaut, 19 Av. Maistriau, 7000 Mons, Belgium. Phone: +32 65 373401, Fax: +32 65 373386, Email: mathieu.ribordy@umh.ac.be} }\\
\vspace*{1.cm}
{\it $^a$ Clark
Atlanta University, Atlanta, GA 30314, U.~S.~A}\\
{\it $^b$University of Mons-Hainaut, 7000 Mons, Belgium}\\

\vspace*{1.0cm}

\end{center}

\medskip\hrule\medskip

\date{June 7, 2005}


\begin{abstract}
{\small 
The event reconstruction in high energy neutrino telescopes is based on the expression of the time delay distribution of the detected photons propagating to some distance from their emission point through a uniform and scattering medium.

Considering a realistic detector with a finite time resolution, we derive an expression for the probability density function of the time delay of detected photons. The asymptotic properties of this function as well as its corresponding cumulative probability distribution are calculated and discussed.
}

\vspace*{0.1cm}
\end{abstract}

\vspace*{1.5cm}
\medskip\hrule\medskip
\section{Introduction}

\setcounter{equation}{0}
\vspace{1mm}

In a scattering medium, the Cherenkov photons emitted along the trajectory of a relativistic charged particle (or resulting from electromagnetic or hadronic showers of neutrino neutral and charged current interactions) and propagating to some distance are delayed relatively to the direct geometrical path.
The expression for the time delay distribution of a photon reaching a detecting device is the essential ingredient in reconstructing events from high energy neutrino telescopes~\cite{exp} and is used to calculate the likelihood of a given event hypothesis.

In this letter, we consider a monochromatic and isotropic point-like emitter-receiver problem: given a uniform medium characterized by scattering and absorbing optical properties and a gaussian time measurement uncertainty, we derive the arrival time delay distribution of the detected photons.
The result is presented in a closed analytical form.

In the following section, we review the photon propagation function in a uniform medium \cite{pandel} and show that it 
satisfies the expected limit of 
a zero time delay when there is neither scattering nor absorption, or when the emitter-receiver distance is zero.

In section ${\bf 3}$, we derive the probability density function (PDF) for a realistic setup, i.e. the detected photons are measured with a finite time resolution. We also discuss the properties of this function, useful in a numerical implementation. 

We conclude this letter with a discussion, presented in section ${\bf 4}$.

\section{Photon propagation function $p$}
\vspace{1mm}

Scattering in a medium causes a photon arrival time delay $t$ defined as a difference between the actual arrival time at the receiver, $t_{arrival}$, and $t_{geom}$, the arrival time from the path of a photon propagating in a medium without scattering:
\begin{equation}
\label{timeresdef1}
t\equiv t_{arrival}-t_{geom}=t_{arrival}-{R\over c}
\end{equation}
where $R$ is the distance between the emitter and the detection point, and $c$ is the light speed in the medium.

Our goal is to determine the PDF describing the distribution of time delay $t$ of photons detected at the distance $R$ from the emitter by the device with a finite time resolution.

Instead of attempting the impregnable task of deriving the PDF from first principles, one might exploit the symmetry of the problem. This suggests the following equation for the photon propagation function $p(R,\,t)$ \cite{pandel}:
\begin{equation}
\label{eq}
p(R_{1}+R_{2},\,t)\,=\,\int^{t}_{0}\,dt^{\prime}\,p(R_{1},\,t^{\prime})\,p(R_{2},\,t-t^{\prime})
\end{equation}
which describes  the picture for $p$, analogous to the Huygens principle for waves in optics.
Eq.~(\ref{eq}) relies on the spherical symmetry of the medium and of the light source. $p(R_{1}+R_{2},\,t)$  is the PDF corresponding to a photon propagation with a time delay $t$ to a distance $R_{1}+R_{2}$ from the source (located at the origin). It can be expressed as the convolution of the PDF to propagate to the shell at distance $R_{1}$ from the origin with the PDF to propagate from this shell up to the shell at distance $R_{1}+R_{2}$ from the origin.
Once multiplied with an exponential damping factor, accounting for absorption and depending on the time delay, the normalized solution of Eq.~(\ref{eq}) can be presented as 
\begin{equation}
\label{pand}
p(\xi,\,\rho,\,t)\,=\, {\rho^{\xi}\,t^{\xi-1}\over \Gamma(\xi)}\,e^{-\,\rho t},\;\;\int^{\infty}_{0}\,dt\,p(\xi,\,\rho,\,t)\,=\,1
\end{equation}
where 
\begin{equation}
\label{parameters}
\xi\equiv {R\over \lambda},\quad \rho\equiv {1\over \tau}\,+\,{c\over \lambda_{a}}
\end{equation}
and $\Gamma(\xi)$ is the gamma function \cite{math}.
In (\ref{parameters}), $\lambda_{a}$ is the absorption length, $\lambda$ and $\tau$ are phenomenological parameters.
When the photon propagation function $p$ is derived from first principles, $\lambda$ and $\tau$ are expressed in terms of the optical parameters of the medium and the differential cross section of the microscopic photon-medium interaction. As they stand in (\ref{pand},\,\ref{parameters}), these parameters must be extracted from the experimental data. 

Clearly, for $R\rightarrow 0$, or for any distance without scattering, there should be no time delay, i.e. $p$ must be concentrated near $t=0$. These two regimes can be realized with the single requirement $\xi\rightarrow 0$, provided that in (\ref{pand},\,\ref{parameters}) $\lambda$ is the effective scattering length in the medium (which is related to the mean free path and to the scattering profile in a way that when the effective scattering length increases, so does the mean free path). 

Examining the case $\xi\rightarrow 0$ shows that $\lim_{\xi\rightarrow 0}p(\xi,\,\rho,\,t)$ does not exist - $p$ is recognized as a generalized function with the kernel $\gamma_{+}(x)\equiv x^{\xi-1}_{+}/\Gamma(\xi)$ \cite{gelfand}. As is well known, generalized functions and their limits are not well defined for any values of their variables and parameters; the self-consistent approach is to consider $(p,\,\varphi)\equiv \int\,p\,\varphi$ instead of $p$, where $\varphi(t)$ is a test function falling off at $t\rightarrow \pm\infty$ fast enough, ensuring the finiteness of $(p,\,\varphi)$ \cite{gelfand}. 

In terms of $(p,\,\varphi)$, the condition $t=0$ for no scattering and/or zero distance can be readily verified: for any analytic function $\varphi(t)$, we have
\begin{equation}
\label{filter}
\lim_{\xi\rightarrow 0}\,(p,\,\varphi)\,=\,\lim_{\xi\rightarrow 0}\int^{\infty}_{0}\,dt\,p(\rho,\,\xi,\,t)\,\varphi(t)\,=\,\lim_{\xi\rightarrow 0}\,\sum^{\infty}_{n=0}\,{\varphi^{(n)}(0)\over n!}\,\int^{\infty}_{0}\,dt\,p(\xi,\,\rho,\,t)\,t^{n}\,=\,\varphi(0)
\end{equation}
meaning that $p$ has the limit (in the sense of generalized functions \cite{gelfand}) 
\begin{equation}
\label{delta}
\lim_{\xi\rightarrow 0}p(\xi,\,\rho,\,t)\,=\,\delta_{+}(t)
\end{equation}
In (\ref{delta}), $\delta_{+}(t)$ is a generalized function, filtering out the point $t=+0$ - origin on the $t$ -axis, approached from the region $t\,>\,0$ (propagation time delay cannot be negative - $p$ is defined on the right semi-axis $t\geq 0$). Eq.~(\ref{delta}) supports our interpretation of $\lambda$ as the effective scattering length in a medium.

The function $p$ describes just the propagation in the medium, and can be interpreted 
as the PDF for the measured photon arrival time assuming an ideal detecting device.

\section{Photon arrival time PDF ${\cal F}_{\sigma}$}
\vspace{1mm}
A PDF for realistic signals should account not only for the photon propagation and 
the properties of the medium, 
but also for the features of the detector.
Here, we consider the finite time resolution of the detector when measuring a signal. The function $p$ described above can be interpreted as a limit corresponding to an ideal measurement. The realistic PDF is thus constructed in terms of $p$ and $\varphi$, the function simulating the detector time resolution distribution.

\begin{figure}[ht]
\vspace{-4mm}
\flushleft
\centering
\begin{minipage}[c]{7.5cm}
\centering
\epsfig{file=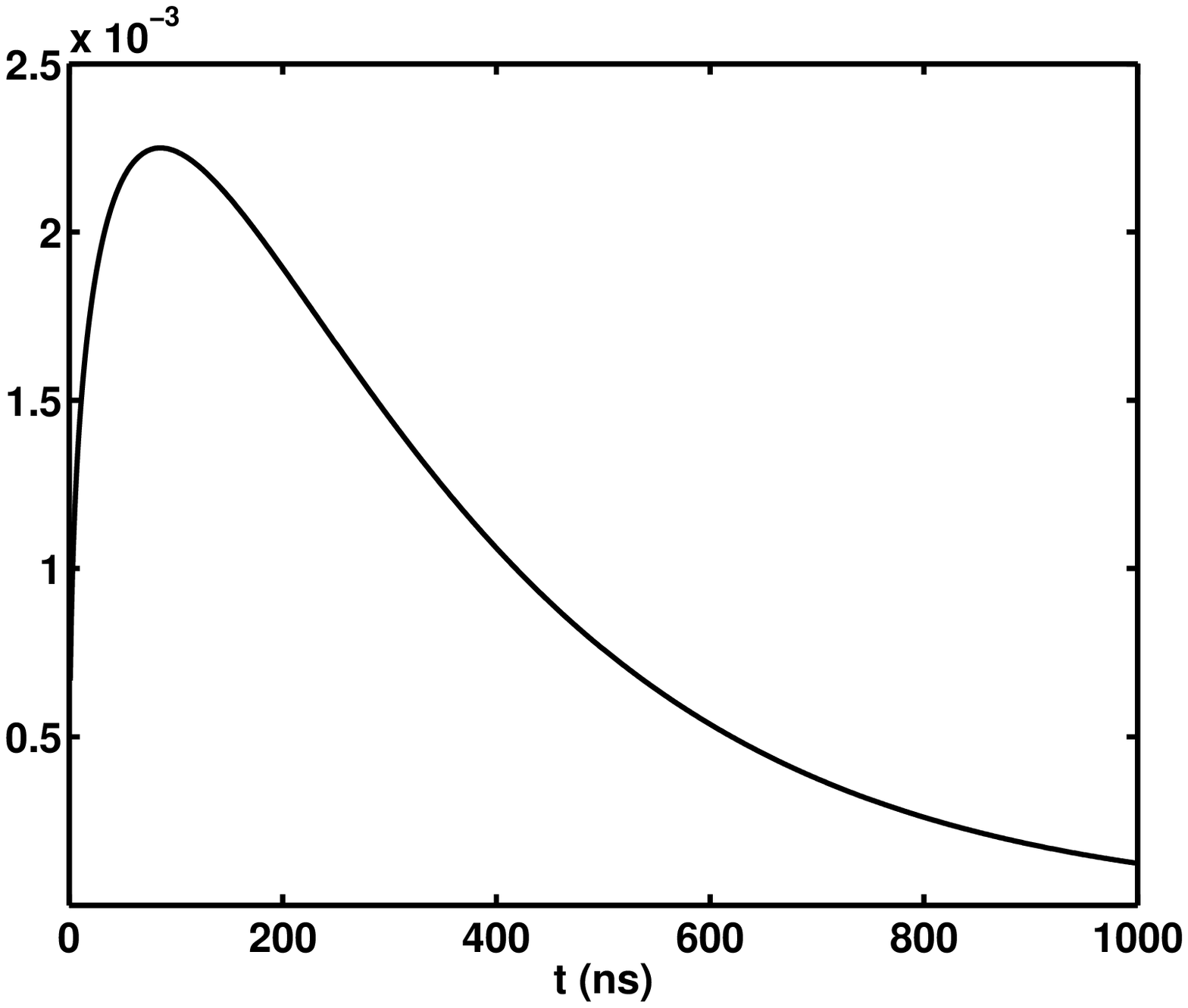,width=7.5cm}
\end{minipage}
\hspace*{0.5cm}
\begin{minipage}[c]{7.5cm}
\centering
\flushleft
\epsfig{file=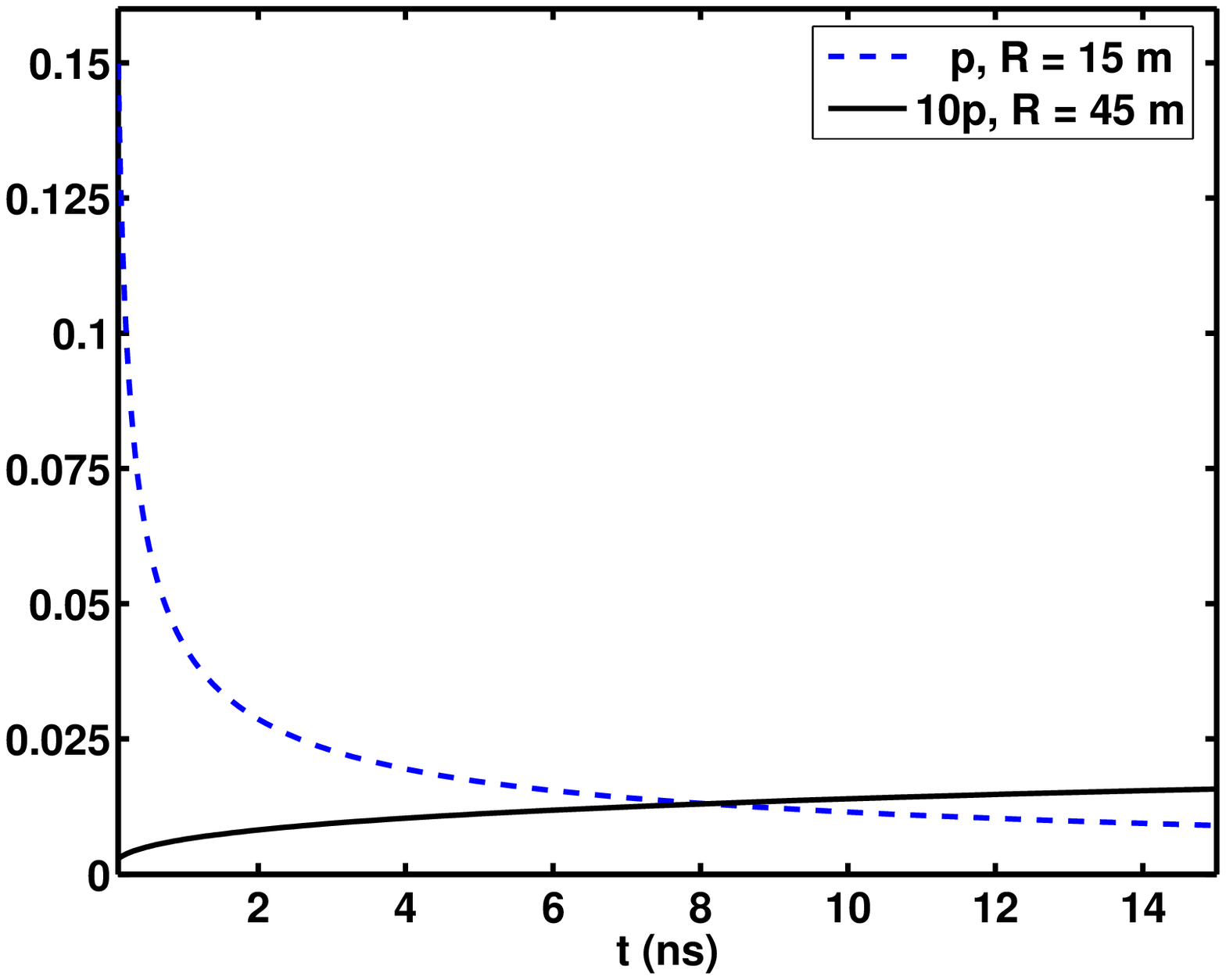,width=7.5cm}
\end{minipage}
\caption{\small
$p$, emitter-receiver distance $R=45\,\,\mathrm{m}\,>\,R_{crit}=\lambda$  (left), and $p$ for $R=15\,\mathrm{m}<R_{crit}=\lambda$, and $10$ times $p$ for $R=45\,\mathrm{m}$ (right). Divergence of $p$ at small $t$ and $R<R_{crit}=\lambda$ is shown. Throughout all the plots the values for the parameters are from  \cite{fit}: $\lambda=33.3$ m, $\lambda_{a}=98$ m and $\tau=557$ ns. 
}
\end{figure}

We choose a gaussian distribution as a ``detector function'' $\varphi(t)$. The width $\sigma\neq0$ of the gaussian distribution represents the time resolution and should be extracted from the characteristics of the detector. $\sigma$ is usually a composite from numerous sources affecting the final detector time resolution (which further motivates the use of a gaussian).

Considering a detector with a finite time resolution instead of an ideal one is satisfactory from the physical point of view. We will see that this also removes the divergence at $t=0$: the PDF accounting for a finite time resolution is an ordinary function which exists for all $R,\,t$.

\subsection{Definition of ${\cal F}_{\sigma}$}
\vspace{1mm}
From the fact that a realistic setup will allow only the observation of the sum of the two independent random variables, the photon time delay and the detector time resolution, it follows that the photon time residual PDF ${\cal F}_{\sigma}$ is the convolution of the photon propagation function $p$ with the detector time jitter function $g_{\sigma}$:
\begin{equation}
\label{convdef}
{\cal F}_{\sigma}(\xi,\,\rho,\,t)\equiv \int^{\infty}_{0}\,dt^{\prime}\,p(\xi,\,\rho,\,t^{\prime})\,g_{\sigma}(t-t^{\prime})
\end{equation}
where
\begin{equation}
\label{gs}
g_{\sigma}(t)\,=\,{\exp(-t^{2}/2\sigma^{2})\over \sqrt{2\pi\sigma^{2}}}
\end{equation}
incorporates the finite time resolution $\sigma\neq 0$.

Since (\ref{convdef}) is of the form $(p,\,\varphi)$, ${\cal F}_{\sigma}$ is not a generalized function, i.e. it exists for any $R$, $t$. 

It should be mentioned that
when the time resolution is accounted for, the definition of the time 
residual $t$ is slightly changed: it does not refer anymore to the photon
arrival time but to the device trigger time - the arrival time corrected with a time jitter. Therefore $t<0$ is acceptable for ${\cal F}_{\sigma}$, in contrast to $p$, where the very definition of the time delay demands to consider only $t\geq 0$.
\begin{figure}[ht]
\vspace{-4mm}
\flushleft
\centering
\begin{minipage}[c]{7.5cm}
\centering
\epsfig{file=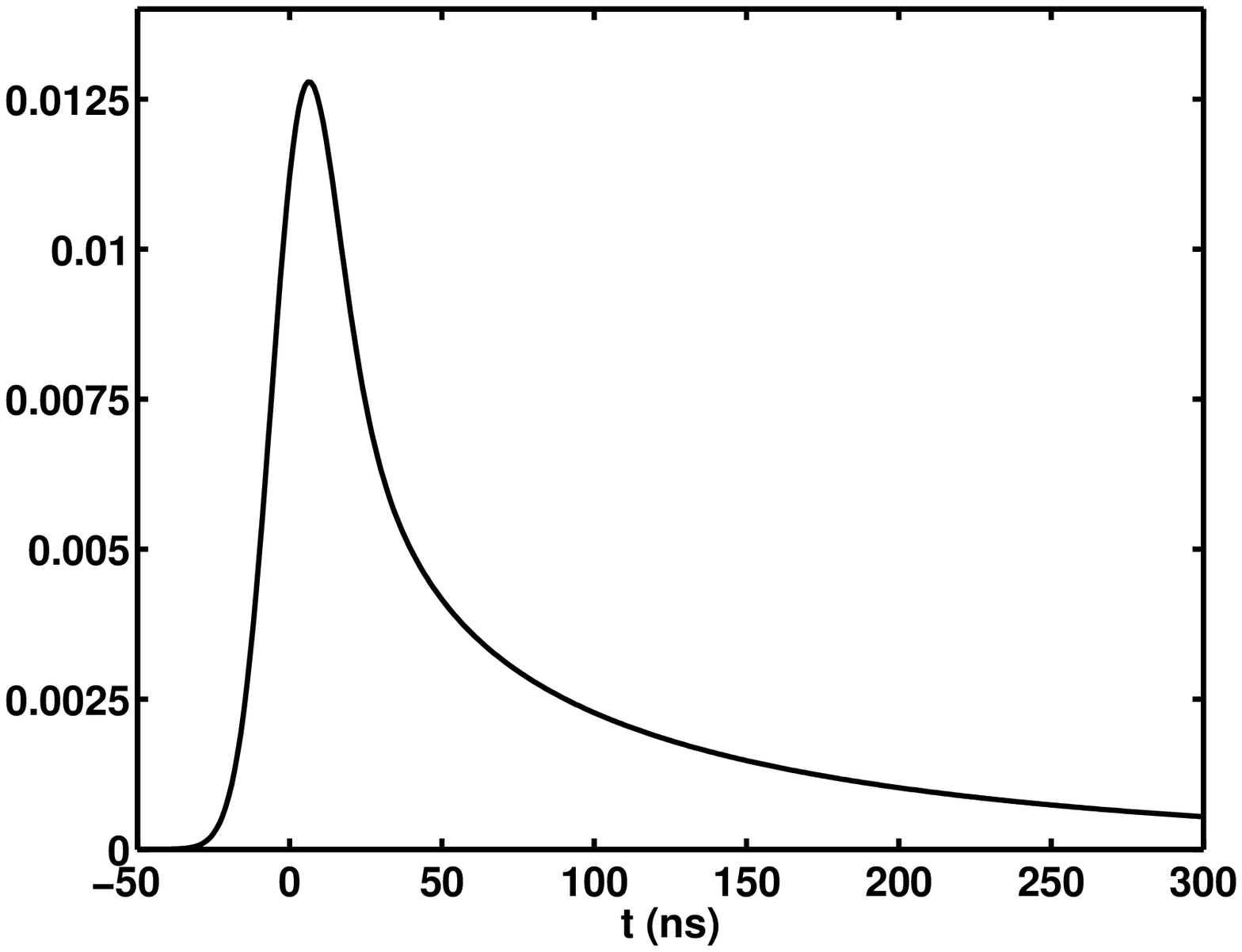,width=7.5cm}
\end{minipage}
\hspace*{0.5cm}
\begin{minipage}[c]{7.5cm}
\centering
\flushleft
\epsfig{file=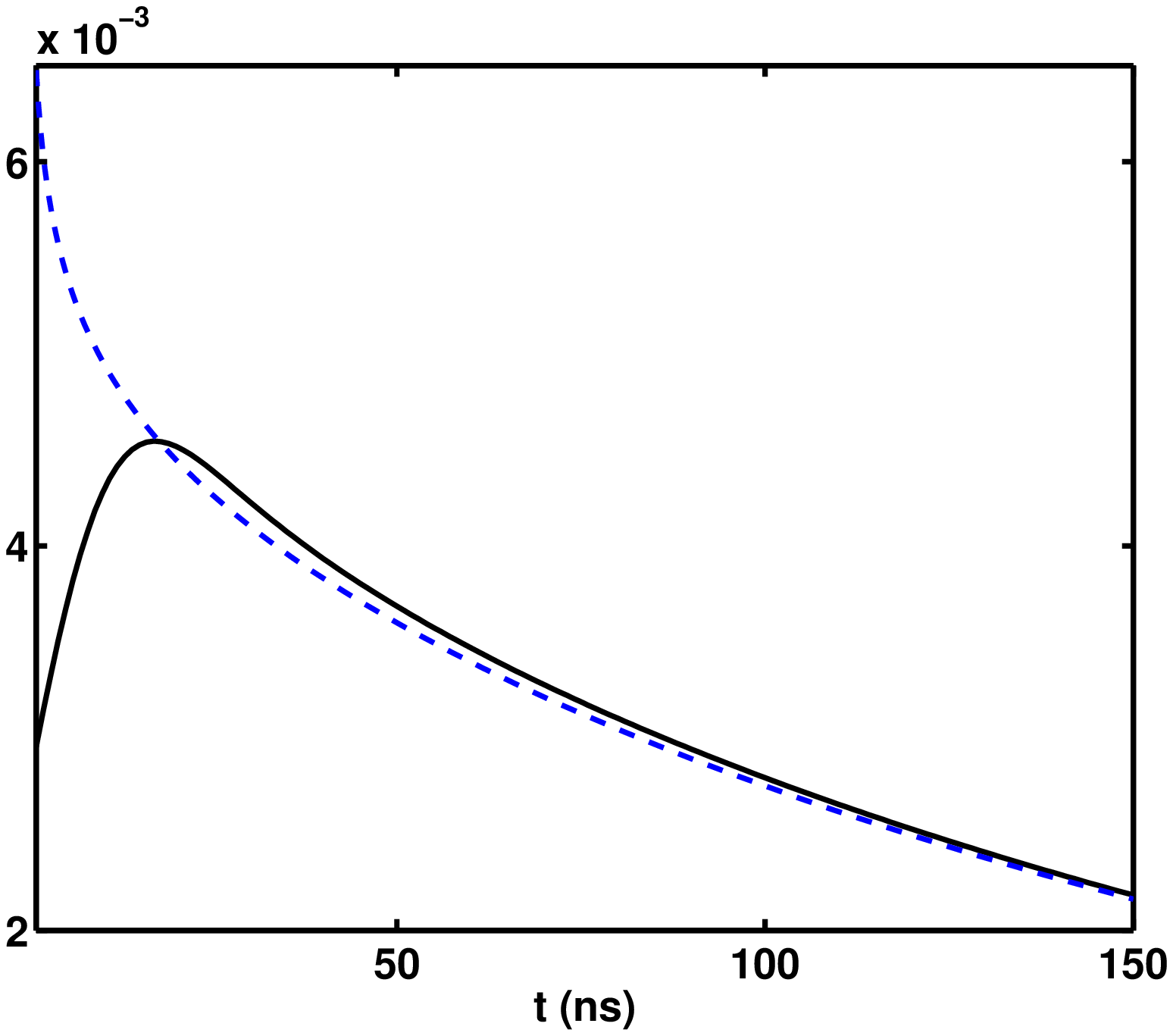,width=7.5cm}
\end{minipage}
\caption{\small
${\cal F}_{\sigma}$, emitter-receiver distance $R=15$ m (left), and ${\cal F}_{\sigma}$ (solid line) and $p$ (dashed line) for the distance $R=0.9\,R_{crit}=0.9\,\lambda=30\,\mathrm{m}$ (right). Time resolution $\sigma=10$ ns. Convolution is finite for any $R\geq 0$, $-\infty\,<\,t\,<\,\infty$, and the spurious singularity at $R\leq \lambda,\,t=0$, present in $p$, is absent in ${\cal F}_{\sigma}$.
}
\end{figure}

Now, the time residual $t$ can also be negative, and the normalization condition
\begin{equation}
\label{norm}
\int^{\infty}_{-\infty}\,dt\,{\cal F}_{\sigma}(\rho,\,\xi,\,t)\,=\,\int^{\infty}_{0}\,dt^{\prime}\,p(\rho,\,\xi,\,t^{\prime})\,=\,1
\end{equation}
is readily verified.

From the physical point of view, negative time residuals do not contradict causality if ${\cal F}_{\sigma}$ decreases faster than $g_{\sigma}$, when $t<0$. In other words, causality is preserved when $g_{\sigma}$ envelops ${\cal F}_{\sigma}$ for negative $t$. 
In the next subsection, we show that this is the case.

The integration in (\ref{convdef}) can be carried out and we obtain:
\begin{equation}
\label{D}
{\cal F}_{\sigma}(\xi,\,\rho,\,t)={(\rho \sigma)^{\xi}\over \sqrt{2\pi\sigma^{2}}}\,\exp\biggl({\rho^{2}\sigma^{2}\over 4}-{\rho t\over 2}-{t^{2}\over 4\sigma^{2}}\biggr)\,D_{-\xi}(\eta)
\end{equation}
where $D$ is the function of parabolic cylinder \cite{math} and 
$$\eta\equiv \rho\,\sigma\,-\,{t / \sigma}$$

For implementation purposes, it is convenient to express ${\cal F}_{\sigma}$ in terms of the confluent hypergeometric function  $_{1}F_{1}$ \cite{math}:
\begin{eqnarray}
{\cal F}_{\sigma}(\xi,\,\rho,\,t) & = & {(\rho \sigma)^{\xi}\,2^{(\xi-3)/2}\over \sqrt{\pi}\,\Gamma(\xi)}{e^{-\,t^{2}/2\sigma^{2}}\over \sigma}
\Biggr[ 
\Gamma\biggl ( {\xi\over 2}\biggr)
\,_{1}F_{1}\biggl ( {\xi\over 2},\,{1\over 2};\,{1\over 2}\,\eta^{2} \biggr )
-\cr
\label{convo}
& - &
\sqrt{2}\,\eta\,
\Gamma\biggl ({\xi\,+\,1\over 2}\biggr )\,_{1}F_{1}\biggl ( {\xi\,+\,1\over 2},\,{3\over 2};\;{1\over 2}\,\eta^{2} \biggr )\Biggr]
\end{eqnarray}

\subsection{Properties and moments of  ${\cal F}_{\sigma}$}
\vspace{1mm}
The confluent hypergeometric function  $_{1}F_{1}$  is implemented in GSL, the GNU Scientific Library \cite{gsl}. 
Here we cite a few useful properties of ${\cal F}_{\sigma}$:
\vspace{6mm}

$\bullet$ Large positive time residuals: $t\gg \sigma+\rho\sigma^{2}$. From the asymptotic for $_{1}F_{1}$ \cite{math}, we obtain:
\begin{eqnarray}
\label{101}
{\cal F}_{\sigma}&=&e^{\rho^{2}\sigma^{2}/2}\,p(\xi,\,\rho,\,t)\,\biggl(1-{\rho\sigma^{2}\over t}\biggr)^{\xi-1}\Biggl[1+{\sigma^{2}(1-\xi)(2-\xi)\over t^{2}}\biggl(1-{\rho\sigma^{2}\over t}\biggr)^{-2}+\cr
& + &{\cal O}\Biggl({\sigma^{4}\over t^{4}}\biggl(1-{\rho\sigma^{2}\over t}\biggr)^{-4}\Biggr)\biggr]  
\end{eqnarray}

$\bullet$ Behavior for negative $t\,<\, 0$. From the expansion of the function of parabolic cylinder $D$ \cite{math}, we obtain for $|t|\,\gg\,\sigma$:
\begin{equation}
\label{cnv2}
{\cal F}_{\sigma}(\rho,\,\xi,\,t)
\approx {(\rho\sigma)^{\xi}\over \sqrt{2\pi\sigma^{2}}}\biggl(\rho\sigma-{t\over \sigma}\biggr)^{-\xi}\exp(-t^{2}/2\sigma^{2})\Biggl[1+{\cal O}\biggl(\rho\sigma-{t\over \sigma}\biggr)^{-1}\Biggr]
\end{equation}
Now, we are in the position to verify that causality is not violated: considering the ratio of ${\cal F}_{\sigma}$ to $g_{\sigma}$, we find that indeed ${\cal F}_{\sigma}$ decreases faster for negative time residual than the detector time jitter function:
\begin{equation}
\label{causality}
{{\cal F}_{\sigma}(t)\over g_{\sigma}(t)}\approx \biggl(1\,-\,{t\over \rho\sigma^{2}}\biggr)^{-\xi}\,\leq\,1
\end{equation}

$\bullet$ $\sigma\,\ll\,1$: using the relation  $\lim_{\sigma\rightarrow 0}\exp(-x^{2}/2\sigma^{2})/ \sqrt{2\pi\sigma^{2}}\,=\,\delta(x)$, we recover $p$ in the zero time resolution limit:
\begin{equation}
\label{2}
\lim_{\sigma\rightarrow 0}{\cal F}_{\sigma}(\xi,\,\rho,\,t)\,=\,p(\xi,\,\rho,\,t)
\end{equation}

$\bullet$ ${\cal F}_{\sigma}$ is finite for $t=0$ and any $R$; e.g. for $R=0$ ($\xi=0$)
\begin{equation}
\label{shortt}
{\cal F}_{\sigma}(0,\,\rho,\,0)\,=\,{1\over \sqrt{2\pi\sigma^{2}}};\quad \lim_{\xi\rightarrow 0,\,\sigma\rightarrow 0}{\cal F}_{\sigma}(\xi,\,\rho,\,t)\,=\,\delta(t)
\end{equation}

$\bullet$ Long distance: for $\xi\,\gg\,1$ we have:
\begin{equation}
\label{long}
{\cal F}_{\sigma}(\xi,\,\rho,\,t)={(\rho\sigma)^{\xi}\xi^{-\xi/2}\over 2\sigma\sqrt{\pi}}\,exp\biggl({\rho^{2}\sigma^{2}\over 4}-{\rho t\over 2}-{t^{2}\over 4\sigma^{2}}+{\xi\over 2}-\eta\sqrt{\xi}\biggr)\,\biggl[1+{\cal O}(\xi^{-1/2})\biggr]
\end{equation}

$\bullet$ Short distance: in the limit $\xi\,\ll\,1$, the receiver time response function survives:
\begin{eqnarray}
\label{3}
{\cal F}_{\sigma}(\xi,\,\rho,\,t)={\exp(-t^{2}/2\sigma^{2})\over \sqrt{2\pi\sigma^{2}}}\,\Biggl[1+{\xi\over 2}\Biggl(2\sqrt{\pi}\ln(\rho\sigma)+\gamma_{E}+2\ln2-\pi \,erfi\biggl({\eta\over \sqrt{2}}\biggr)\Biggr)+{\cal O}(\xi^{2})\Biggr]
\end{eqnarray}
where $erfi(z)$ is an error function of the imaginary argument, $erfi(z)\,\equiv\,{2\over \sqrt{\pi}}\,\int^{z}_{0}\,dx\,e^{x^{2}}$, and $\gamma_{E}\approx 0.57721$ is the Euler's constant \cite{math}.

These properties are useful to implement algorithms and to avoid numerical overflows. E.g., for large $t$, the calculation speed, as implemented in GSL~\cite{gsl}, of $_{1}F_{1}(a,\,b;\,x)$ slows down with increasing $x$ and numerical overflow may occur. This can be circumvented by switching to the asymptotic behavior of ${\cal F}_{\sigma}$ for large $t$. The case of large $\xi$ can be treated using (\ref{long}) to avoid numerical overflow caused by $\Gamma(\xi)$, appearing in (\ref{convo}) and in the expression for the cumulative of ${\cal F}_{\sigma}$ (see below).

Also interesting are the moments of ${\cal F}_{\sigma}$:
\begin{equation}
\label{momdef}
\langle t^{N} \rangle\,\equiv\,\int^{\infty}_{-\infty}\,dt\,t^{N}\,{\cal F}_{\sigma}(\rho,\,\xi,\,t)
\end{equation}
which can be calculated in a closed analytical form; we explicitly present here the first two - the time residual average and the corresponding dispersion:
\begin{equation}
\label{moments}
\langle\,t\,\rangle\,=\,{\xi\over \rho},\quad \langle (\triangle t)^{2}\rangle\equiv\langle\,t^{2}\,\rangle\,-\,\langle\,t\,\rangle^{2}\,=\,\sigma^{2}\,+\,{\xi\over \rho^{2}}
\end{equation}
As expected, only the detector time resolution survives in the variance at $\xi=0$.

\subsection{Cumulative of ${\cal F}_{\sigma}$}
\vspace{1mm}
${\cal F}_{\sigma}$ is a PDF for a single detected photon. Another object of interest is the time delay probability density of the first photon in a sequence of $N$ detected photons. It is given by 
\begin{equation}
\label{Nphotons}
{\cal F}_{\sigma}^{(N)}(\rho,\,\xi,\,t) = N\cdot {\cal F}_{\sigma}(\rho,\,\xi,\,t) \cdot \biggl(1\,-\,{\cal C}_{{\cal F}}(\rho,\,\xi,\,t) \biggr)^{N-1}
\end{equation}
where ${\cal C}_{{\cal F}}$ is the cumulative probability: 
\begin{equation}
\label{cumdef}
{\cal C}_{{\cal F}}(\rho,\,\xi,\,t)  \equiv  \int^{t}_{-\infty}\,dt\,{\cal F}_{\sigma}(\rho,\,\xi,\,t)
\end{equation}

This function appears in the expression for the probability $\omega(\rho,\,\xi,\,\sigma,\,t;\,\epsilon)$ that the time residual lies between $t$ and $t\,+\,\epsilon$:
\begin{equation}
\label{prob}
\omega(\rho,\,\xi,\,\sigma,\,t;\,\epsilon)\equiv \int^{t+\epsilon}_{t}\,dt^{\prime}\,{\cal F}_{\sigma}(\rho,\xi,t^{\prime})\,=\,{\cal C}_{{\cal F}}(\rho,\,\xi,\,t+\epsilon)\,-\,{\cal C}_{{\cal F}}(\rho,\,\xi,\,t) 
\end{equation}
Note that since  ${\cal F}_{\sigma}$ is a smooth function defined everywhere, for $\epsilon\ll t$ the probability above can be approximated by ${\cal F}_{\sigma}$:
\begin{equation}
\label{appr}
\omega(\rho,\,\xi,\,\sigma,\,t;\,\epsilon)\approx \epsilon\,{\cal F}_{\sigma}(\rho,\,\xi,\,t+\epsilon/2)
\end{equation}
This prescription cannot be used in the case $\sigma=0$. Instead, the analytical solution for the cumulative distribution ${\cal C}_{p}$, given in terms of incomplete gamma function \cite{math} $\Gamma(\xi,\,\rho t)$, must be used: $${\cal C}_{p}\equiv\int^{t}_{0}\,dt^{\prime}\,p(\rho,\,\xi,\,t^{\prime})=1-\Gamma(\xi,\,\rho t)/\Gamma(\xi)$$

The analytical expression for ${\cal C}_{{\cal F}}$ with an arbitrary $\xi$ cannot be obtained.
\begin{figure}
\vspace*{-7mm}
\centering
\epsfig{file=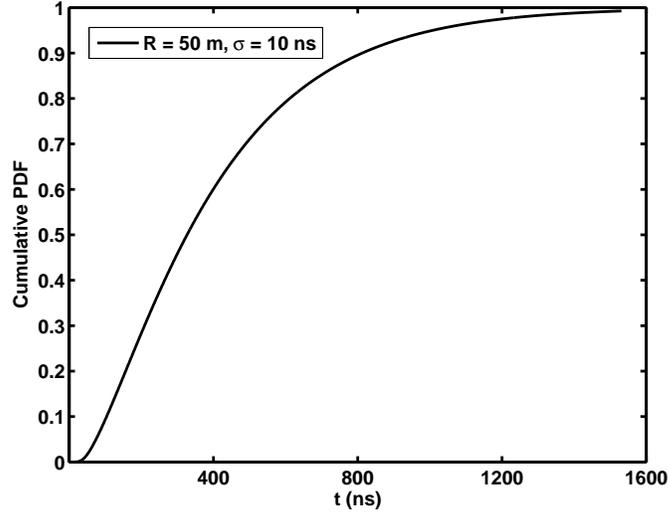,width=9.9cm}
\caption{\small
Cumulative PDF ${\cal C}_{{\cal F}}$ versus $t$ (ns) for $R=50$ m. Time resolution $\sigma=10$ ns.
}
\end{figure}
From  (\ref{convo}), it follows that the cumulative distribution can be expressed as an integral involving the error function:
\begin{eqnarray}
\label{cumu}
{\cal C}_{{\cal F}}(\rho,\,\xi,\,t)\,=\,{1\over 2}\,+\,{\rho^{\xi}\over 2\Gamma(\xi)}\,\int^{\infty}_{0}\,dy\,y^{\xi-1}\,e^{-\rho y}\,erf\biggl({t-y\over \sqrt{2\sigma^{2}}}\biggr)
\end{eqnarray}
When $\xi\,=\,1$ the result is
\begin{equation}
\label{xi1}
{\cal C}_{{\cal F}}(\rho,\xi,\,t)|_{\xi=1}\,={1\over 2}\,+\,{1\over 2}\,\Biggl[erf\biggl({t\over \sqrt{2\sigma^{2}}}\biggr)\,-\,\exp(\rho^{2}\sigma^{2}/2\,-\,\rho t)\,\Biggl(1\,-\,erf\biggl(\rho\sqrt{{\sigma^{2}\over 2}}\,-\,{t\over \sqrt{2 \sigma^{2}}}\biggr)\Biggr)\Biggr]
\end{equation}
The case of integer $\xi$ can be handled using
\begin{equation}
\label{xiint}
\int^{\infty}_{0}dyy^{n}e^{-\rho y}erf(cy\,+\,b)=(-1)^{n}{\partial^{n}\over\partial \rho^{n}}{1\over \rho}\,\Biggl[erf(b)\,+\,\exp((\rho^{2}+4bc)/4c^{2})\,\Biggl(1\,-\,erf\biggl(b\,+{\rho\over 2c}\biggr)\Biggr)\Biggr]
\end{equation}

An analytic continuation for a fractional order derivative is possible \cite{gelfand}, but not practical as higher hypergeometric functions emerge. In the Appendix, we give a simple semi-numerical approximation valid for any $\xi\geq 0$.

In \cite{fit}, further examples involving the cumulative distribution are presented.

\section{Discussion}
In this letter, we suggested and calculated the arrival time probability density for a photon at some distance from an isotropic point-like source in a uniform scattering medium for a realistic detector. ${\cal F}_{\sigma}$, accounting for a finite detector time resolution $\sigma\neq0$, is free from the spurious singularities which appear in the generalized function $p$ (the unrealistic zero time resolution limit of ${\cal F}_{\sigma}$) and is defined for all $t$ and $R\geq0$. The expression for ${\cal F}_{\sigma}$ has been obtained in a closed analytic form ({\ref{D},\,\ref{convo}) and contains 
parameters characterizing the medium and the detector.
We further discussed properties of ${\cal F}_{\sigma}$, useful for a numerical implementation. 
Finally, we calculated semi-analytically the corresponding cumulative probability entering multi-photon PDF's in order to provide a usable numerical implementation scheme.  
\\[8mm]

{\large{\bf Acknowledgments.}} This research was supported by the National Science Foundation (NSF-G067771) and the Swiss National Foundation (PA002-104970). We would like to thank A.~Bouchta, T.~Castermans, F.~Grard, P.~Herquet, R.~Porrata and C.~Wiebusch for valuable comments.

\section*{Appendix: Numerical integration scheme for ${\cal C}_{{\cal F}}$}
\vspace{1mm}
We sketch the procedure for the calculation of the cumulative ${\cal C}_{{\cal F}}$.

To avoid numerical integration from $0$ to $\infty$ in (\ref{cumu}), we use the asymptotic properties of the error function \cite{math}, and split the integral on three subintervals: $(0,\,\infty)=(0,\,y_{\mathrm{min}})\oplus(y_{\mathrm{min}},\,y_{\mathrm{max}})\oplus(y_{\mathrm{max}},\,\infty)$. The parameter $\alpha$ in $y_{\mathrm{min}} =\max{(0,t\,-\,\alpha\,\sqrt{2\sigma^2})}$ and
$y_{\mathrm{max}}=\max{(0,t\,+\,\alpha\,\sqrt{2\sigma^2})}$ is chosen 
from the condition ${|t\,-\,y|\over \sqrt{2\sigma^2}}>\alpha$, leading to 
$erf((t\,-\,y)/ \sqrt{2\sigma^2})\approx \pm1$. E.g., choosing $\alpha\,=\,2.5$ results in an error less than $0.041$\%.

The integration on the first and third subintervals, where $erf=\pm1$, can be carried out and we are left with an integration over the finite interval $(y_{\mathrm{min}},\,y_{\mathrm{max}})$:
$$
{\cal C}_{{\cal F}}(\rho,\,\xi,\,t)=
1-\frac{\Gamma(\xi,\rho y_{\mathrm{max}})+\Gamma(\xi,\rho y_{\mathrm{min}})}{2\Gamma(\xi)}
+J;\quad J\,\equiv\,{\rho^{\xi}\over 2\Gamma(\xi)}\,
\int^{y_{\mathrm{max}}}_{y_{\mathrm{min}}}\,dy\,y^{\xi-1}\,e^{-\rho y}\,erf\Biggl( {t\,-\,y\over \sqrt{2\sigma^2}}\Biggr)
$$
where $\Gamma(a,\,x)\equiv \int^{\infty}_{x}\,dz\,z^{a-1}e^{-z}$, the incomplete gamma function \cite{math}, is implemented in GSL \cite{gsl}.

Integral with finite limits $J$ is finite and can in principle be calculated numerically, but when $\xi\,<\,1$, numerical integration algorithms may become unstable when $y_{\mathrm{min}}\rightarrow 0$. This can be circumvented by handling $J$ separately for the two cases:
\begin{enumerate} 
\item $\xi\,>\, 1$. 

$J$ is calculated numerically.  
Since $y_{\mathrm{min}}$ is positive and $\xi\,>\,1$, there is no problem with the convergence; in this case, a standard integration algorithm \cite{c++} can be applied.
\item $\xi\,<1\,$. 

When $y_{\mathrm{min}}\,\ll\,1$, $y^{\xi}$ is an integrable singularity in the region $y\sim y_{\mathrm{min}}$.
In order to isolate explicitly the terms $\sim y^{\xi}_{\mathrm{min}}$, a small finite $\varepsilon$,  $y_{\mathrm{min}}\,<\,\varepsilon\,<\,y_{\mathrm{max}}$, is introduced and the integration is performed analytically in the range $y_{\mathrm{min}}\,<\,y\,<\varepsilon$.
Expanding $\exp(-\rho y)$ and $erf((t-y)/\sqrt{2\sigma^{2}})$ in $y$, we obtain:
\begin{eqnarray}
\nonumber
J & = & \,{\rho^{\xi}\over 2\Gamma(\xi)}\,\Biggl[\int^{\epsilon}_{y_{\mathrm{min}}}\,+\,\int^{y_{\mathrm{max}}}_{\varepsilon}\Biggr]\,dy\,y^{\xi-1}\,e^{-\rho y}\,erf\Biggl( {t\,-\,y\over \sqrt{2\sigma^2}}\Biggr)\,=\cr
& = & {\rho^{\xi}\over 2\Gamma(\xi)}\,\Biggl[\,\int^{y_{\mathrm{max}}}_{\varepsilon}\,dy\,y^{\xi-1}\,e^{-\rho y}\,erf\Biggl( {t\,-\,y\over \sqrt{2\sigma^2}}\Biggr)\,+\cr
\nonumber
& + &\int^{\epsilon}_{y_{\mathrm{min}}}\,dy\,y^{\xi-1}\,(1-\rho y)\,\Biggl(erf\Biggl( {t\over \sqrt{2\sigma^2}}\Biggr)\,-\,y\,e^{-t^{2}/2\sigma^{2}}\sqrt{2\over \pi\sigma^{2}}\Biggr)\Biggr]\,=\cr
\nonumber
& = & {\rho^{\xi}\over 2\Gamma(\xi)}\,\Biggl[\,\int^{y_{\mathrm{max}}}_{\varepsilon}\,dy\,y^{\xi-1}\,e^{-\rho y}\,erf\Biggl( {t\,-\,y\over \sqrt{2\sigma^2}}\Biggr)\,+\cr
\nonumber
& + & erf\Biggl( {t\over \sqrt{2\sigma^2}}\Biggr)\,{\varepsilon^{\xi}-y^{\xi}_{\mathrm{min}}\over \xi}\,-\,\Biggl(\sqrt{2\over \pi\sigma^{2}}\,e^{-t^{2}/2\sigma^{2}}\,+\cr
\label{WWW}
& + &\rho\,erf\Biggl( {t\over \sqrt{2\sigma^2}}\Biggr)\Biggr)\,{\varepsilon^{\xi+1}-  y^{\xi+1}_{\mathrm{min}}\over \xi+1}\,+\,O(\varepsilon^{\xi+2})\Biggr]
\end{eqnarray}
and the remaining integral
$$
\int^{y_{\mathrm{max}}}_{\varepsilon}\,dy\,y^{\xi-1}\,e^{-\rho y}\,erf( (t\,-\,y)/ \sqrt{2\sigma^2})
$$
can be calculated numerically using standard integration algorithms.
\end{enumerate}


\begin{thebibliography}{**}
\bibitem{exp}
Baikal - http://www.ifh.de/baikal/baikalhome.html; ANTARES - http://antares.in2p3.fr\\ AMANDA - http://amanda.wisc.edu;  IceCube - http://icecube.wisc.edu
\bibitem{pandel}
D.~Pandel, Bestimmung von Wasser- und Detektorparametern und Rekonstruktion von Myonen bis 100 TeV mit dem Baikal-Neutrinoteleskop NT-72, Diploma Thesis, Humboldt-Universit\"{a}t zu Berlin, Berlin, Germany (Feb.  1996).
\bibitem{math}
M.~Abramowitz and I.~A.~Stegun, ``Handbook of Mathematical Functions'', NY: Dover, (1972).
\bibitem{gelfand}
I.~M.~Gelfand and G.~Shilov, ``Generalized Functions'', NY: Academic Press,  (1968).
\bibitem{fit}
J.~Ahrens at et al, Muon Track reconstruction and Data Selection Techniques in AMANDA, Nuclear Instruments \& Methods A524, 169-194, (2004).
\bibitem{gsl}
http://sources.redhat.com/gsl
\bibitem{c++}
W.~H.~Press, S.~A.~Teukolsky, W.~V.~Vetterling, B.~P.~Flannery, ``Numerical Recipes in C++ - The Art of Scientific Computing'',  Cambridge University Press, Cambridge, UK, 2002.
\end{thebibliography}
\end{document}